\documentclass{article}

\usepackage[hmarginratio=1:1,top=32mm,columnsep=20pt]{geometry}
\usepackage{amsmath,amssymb,amstext,amsthm}
\usepackage[square,colon,authoryear]{natbib}
\usepackage[pdftex]{graphicx}
\usepackage{hyperref,color}
\usepackage{lmodern}
\usepackage{titlesec}
\usepackage{multirow}

\title{\textbf{A New Method for Performance
Analysis \\ in Nonlinear Dimensionality Reduction}}

\author{
\large
\textsc{Jiaxi Liang, Shojaeddin Chenouri \thanks{Corresponding author: schenouri@uwaterloo.ca} and Christopher G. Small} \\[2mm]
\\
 \normalsize Department of Statistics \& Actuarial Science\\ 
  \normalsize University of Waterloo \\
\vspace{-5mm}
}
\date{}

\begin{document}

\maketitle % Insert title
%----------------------------------------------------------------------------------------
%	ABSTRACT
%----------------------------------------------------------------------------------------

\begin{abstract}
In this paper, we develop a local rank correlation measure which quantifies the performance of dimension reduction methods.  
The local rank correlation is easily interpretable, and robust against the extreme skewness of nearest neighbor distributions in high dimensions. Some benchmark datasets are studied. We find that the local rank correlation closely corresponds to our visual interpretation of the quality of the output.
In addition, we demonstrate that the local rank correlation is useful in estimating the intrinsic dimensionality of the original data, and in selecting a suitable value of tuning parameters used in some algorithms.
\end{abstract}
KEYWORDS: Rank correlation, Manifold, Dimension reduction, Isomap, Local tangent space alignment, Maximum variance unfolding, Principal component analysis.
\maketitle

\section{Introduction}

With recent advances in data collection and storage capabilities, researchers across a wide variety of fields are facing larger datasets with increasing dimensionality, arising from images, videos, text documents, etc. Dimensionality reduction methods have been developed and applied as pre-processing tools to deal with such high-dimensional datasets. 

In dimensionality reduction, we assume that the observed data in high-dimensional space lie on (or near) an embedded submanifold with lower dimensionality. With this fundamental assumption, it is often possible to represent the high-dimensional data in a low-dimensional space without losing important information from the original data. 

Unless stated otherwise, in this paper all vectors are assumed to be column vectors. For a vector $\mathbf{x}$, its transpose is denoted by $\mathbf{x}^{\prime}$. Formally, we state the dimension reduction problem as follows. Suppose that there are $n$ points in a $q$-dimensional space $\mathbb{R}^{q}$, denoted individually by $\left\{ \mathbf{y}_{1},\ldots,\mathbf{y}_{n} \right\}$, or collectively by an $n\times q$ matrix $ \mathbf{Y} $ with the $j$-th row being $\mathbf{y}_{j}^{\prime}$. Furthermore, we assume $\mathbf{y}_{1},\ldots,\mathbf{y}_{n}$ are mapped into a higher-dimensional space $\mathbb{R}^{p}$ by an unknown isometry $\varphi:\mathbb{R}^{q} \rightarrow \mathcal{M}^{q} \subset \mathbb{R}^{p}$ ($p>q$) possibly with added random error: 

\begin{equation}
\label{eq1}
\mathbf{x}_{j}=\varphi(\mathbf{y}_{j})+\pmb{\epsilon}_{j}\,,
\end{equation}
where $\pmb{\epsilon}_{j}\in \mathbb{R}^{p},j=1,\ldots,n$ are i.i.d. with mean 0, and $\mathcal{M}^{q}$ is a differentiable embedded submanifold.

We represent $\left\{ \mathbf{x}_{1},\ldots,\mathbf{x}_{n} \right\}$ collectively as an $n\times p$ matrix $ \mathbf{X} $. The vectors $\left\{ \mathbf{y}_{1},\ldots,\mathbf{y}_{n} \right\}$ are not observed, and we only observe $\left\{ \mathbf{x}_{1},\ldots,\mathbf{x}_{n} \right\}$. We say $\left\{ \mathbf{x}_{1},\ldots,\mathbf{x}_{n} \right\}$ lie on (or near) the manifold $\mathcal{M}^{q}$ with \textsl{intrinsic dimensionality} $q$, or we say the intrinsic dimensionality of $\mathbf{X}$ is $q$. By ``case $j$'' we shall mean the index of the corresponding points $\mathbf{x}_{j}$ and $\mathbf{y}_{j}$.

The purpose of dimensionality reduction algorithms is to reconstruct the inverse mapping $\psi=\varphi^{-1}$ up to an arbitrary isometry, and to reconstruct $\mathbf{y}$ by $ \widehat{\mathbf{y}}=\widehat{\psi}(\mathbf{x}) $, again up to an arbitrary isometry. Some dimension reduction methods do not reconstruct $\psi$ directly, but just $\mathbf{y}$. We denote a given dimensionality reduction method as a mapping $\widehat{\psi}:\mathbb{R}^{p}\rightarrow \mathbb{R}^{q}$, and we also write the low-dimensional configuration $\widehat{\mathbf{Y}}$ as $\widehat{\mathbf{Y}}=\widehat{\psi}(\mathbf{X})$ for convenience.

Dimension reduction algorithms can be broadly classified into two types: linear and nonlinear, depending on whether $\widehat{\psi}$ is a linear mapping or not. When $\mathcal{M}$ is assumed to be a $q$-flat in $\mathbb{R}^{p}$, it is natural to use a linear method. Well-known linear techniques include principal component analysis \citep{pca1} and metric multidimensional scaling \citep{mds1}. Nonlinear dimension reduction methods cover all other cases, and they will typically be used when $\mathcal{M}$ is assumed to have some extrinsic curvature, which means $\mathcal{M}$ is not a flat. Over the last two decades, many nonlinear methods have been introduced in the literature. These include Kernel PCA \citep{kpca}, Local Linear Embedding (LLE) \citep{lle}, Isometric Feature Mapping (Isomap) \citep{isomap}, Laplacian Eigenmap \citep{LEM}, Maximum Variance Unfolding (MVU) \citep{mvu} and Local Tangent Space Alignment (LTSA) \citep{ltsa}, among others. Summaries
and surveys can be found in many books and papers \citep{sum1,sum2,sum3}.

There is no universal agreement on how to assess and compare the performance of these different methods. In the supervised learning problems, such as regression or classification, a performance measure can be defined on the response variable (such as prediction error or classification error). But dimensionality reduction, as we formulated here, is an unsupervised learning problem without such a measure. Therefore, a performance measure is needed to quantify the quality of different dimension reduction algorithms. Ideally, we expect this 
\begin{itemize}
\item to be easily interpretable,

\item to be applicable to most algorithms and datasets,

\item to have the property that any tuning parameters of the measure can be easily selected by a simple method.
\end{itemize} 
In addition to these broad criteria, other considerations are necessary.  For example, data which arise in high dimensional contexts will often have interpoint distance distributions which are highly skewed, even locally for nearest neighbor distances.  This is part of the so-called ``curse of dimensionality.''  When this occurs, standard multivariate methods, such as arise from Gaussian assumptions used in low dimensions, are not suitable.  Distances with heavily skewed distributions can be transformed to more symmetric variables on an ad hoc basis.  An alternative, which we choose here, is to use ranking methods for nearest neighbor distances.  Therefore we can add the following requirement.
\begin{itemize}
\item We expect this method to avoid the problems associated with the curse of dimensionality.
\end{itemize}
In this paper, we develop a local rank-correlation measure to accomplish these goals. 

In equation \eqref{eq1}, we have assumed that $\varphi$ is an isometry. It follows that the neighboring points in the input space should be mapped to neighbors in the output space, and vice versa for the inverse mapping $\psi$. This can be called ``neigborhood preservation.'' 
Early attempts to quantify the neigborhood preservation of a dimension reduction method were made in the study of Self-Organizing Maps \citep{SOM}. where measures such as the \textit{topographic product} \citep{qa1}, \textit{topographic function} \citep{qa2} and \textit{quantization error} \citep{qa6} were developed. Advantages and disadvantages of each method are discussed in detail by P\"{o}lzlbauer \citep{qa5}.

More recently, a few rank-based measures have been proposed. These include \textit{mean relative rank errors} (MRREs) \citep{rank2}, \textit{trustworthiness and continuity} (T\&C) \citep{rank1}, \textit{local continuity meta criterion} (LCMC) \citep{rank5}, and the \textit{agreement rate metric} (AR) \citep{rank4}. 
Measures MRREs and T\&C are restricted to the interval $[0,\,1]$ and order outcomes of algorithms in such a way that larger values of these measures are desirable. Both of these measures try to quantify two types of errors that occur during the dimension reduction procedures,
\begin{itemize}
\item[(i)] non-neighboring points in $\mathbb{R}^{p}$ are mapped by $\widehat{\psi}$ to be neighboring points in $\mathbb{R}^{q}$,

\item[(ii)] neighboring points in $\mathbb{R}^{p}$ are mapped by $\widehat{\psi}$ to be non-neighboring points in $\mathbb{R}^{q}$.
\end{itemize}

These two types of errors create a discrepancy between nearest neighbor ranks in the input and output spaces. Therefore they can be measured by calculating the change of nearest neighbor ranks. 
Measure ${\rm AR}_{_J}$ is the average size of the overlap of $J$-nearest neighborhoods between the low-dimensional reconstruction and the original data, while ${\rm LCMC}_{_J}$ accounts for the expected random overlap. Besides the neighborhood preservation measures, \cite{PA} proposed a Procrustes measure that evaluates how well each local neighborhood matches its corresponding embedding under an optimal linear transformation.  

The rest of the paper is organized as follows. In Section 2, we introduce a new class of local rank correlation measures. In Section 3, we provide some applications on benchmark datasets. Finally, in Sections 4 and 5, we employ local rank correlations to choose suitable values of parameters used in modelling and tuning. 

\section{Local rank correlation}

For an observed high-dimensional dataset $ \left\{\mathbf{x}_{1},\ldots,\mathbf{x}_{n} \right\} \subset \mathcal{M}$ and a low-dimensional representation $\left\{ \widehat{\mathbf{y}}_{1},\ldots,\widehat{\mathbf{y}}_{n} \right\}$, we have the following notation:

\begin{itemize}
\item $\left\| \cdot \right\|$: the Euclidean norm.

$d^{\mathcal{M}}(\mathbf{x}_1,\,\mathbf{x}_2)$: the geodesic distance from $\mathbf{x}_1$ to $\mathbf{x}_2$ on the Riemannian manifold $\mathcal{M}$.

\bigskip

\item $s_{_{ij}}$: the rank of $\left\|\mathbf{x}_{i}-\mathbf{x}_{j}\right\|$ in ascending order, counting outward from $\mathbf{x}_{i}$, that is 
$$ s_{_{ij}}=\#\left\{1\le k\le n\,\,:\,\left\|\mathbf{x}_{i}-\mathbf{x}_{k}\right\|\leq\left\|\mathbf{x}_{i}-\mathbf{x}_{j}\right\|\right\}$$

$r_{ij}$: the rank of $\left\|\mathbf{y}_{i}-\mathbf{y}_{j}\right\|$ in ascending order, counting outward from $\mathbf{y}_{i}$, that is 
$$ r_{_{ij}}=\#\left\{1\le k\le n\,\,:\,\left\|\mathbf{y}_{i}-\mathbf{y}_{k}\right\|\leq\left\|\mathbf{y}_{i}-\mathbf{y}_{j}\right\|\right\}$$

$\widehat{r}_{_{ij}}$: the rank of $\left\|\widehat{\mathbf{y}}_{i}-\widehat{\mathbf{y}}_{j}\right\|$ in ascending order, counting outward from $\widehat{\mathbf{y}}_{i}$, that is 
$$ r_{_{ij}}=\#\left\{1\le k\le n\,\,:\,\left\|\widehat{\mathbf{y}}_{i}-\widehat{\mathbf{y}}_{k}\right\|\leq\left\|\widehat{\mathbf{y}}_{i}-\widehat{\mathbf{y}}_{j}\right\|\right\}$$

\bigskip

\item $N^{^I}_{_J}(i)$: the index set of $J$-nearest neighbors of $\mathbf{x}_{i}$, 
that is $N^{^I}_{_J}(i)=\left\{ j\,|\,1 \leq s_{_{ij}}\leq J \right\}$.

$N^{^O}_{_J}(i)$: the index set of $J$-nearest neighbors of $\mathbf{y}_{i}$, that is 
$N^{^O}_{_J}(i)=\left\{ j|1 \leq \widehat{r}_{_{ij}}\leq J \right\}$. 

Here the superscripts I/O stand for input/output spaces of the algorithms.
\bigskip

\item $\# A$: the cardinality of the set $A$.
\end{itemize}

In the problem setup \eqref{eq1}, where $\varphi$ is assumed to be an isometry, we can suppose that geodesic distances are well approximated by Euclidean distances on the local level.  Therefore, we might ideally assume that for any point $\mathbf{x}_{i}$, there exists a set $N(i)$ of neighboring cases such that the nearest neighbor ranks of the latent low-dimensional data $\mathbf{Y}$ are preserved in $\mathbf{X}$, i.e.
$$r_{_{ij}}=s_{_{ij}}.$$

Therefore, a low-dimensional representation $\widehat{\mathbf{Y}}$ can be said to have rank fidelity if $\widehat{\mathbf{Y}}$ also preserves such ranks, i.e. 
$$\widehat{r}_{_{ij}}=s_{_{ij}}.$$
This identity assumes that there is a bijection between the data points in the input and output neighborhoods. In practice when perfect rank fidelity is not achieved then no bijection can be assumed. Therefore two types of errors could occur due to the mapping $\widehat{\psi}$.
\begin{itemize}
\item Output error: The changes of nearest neighbor ranks $\widehat{r}_{_{ij}}$ from the output space to the input space.

\item Input error: The changes of nearest neighbor ranks $s_{_{ij}}$ from the input space to the output space.
\end{itemize}

These two types of errors can be measured by a local rank correlation between the nearest neighbor distances in the input and output spaces. For a given $i$, and all $j$ in $N_{J}^{I}(i) \bigcup N_{J}^{O}(i)$, define the trimmed rank
\begin{align}
\label{ad1}
& S_{_{ij}}=
\begin{cases}
\delta_{_{ij}}, & j \in N_{J}^{I}(i) \bigcap N_{J}^{O}(i)\\
\frac{\zeta+J+1}{2}, & j \notin N_{J}^{I}(i) \bigcap N_{J}^{O}(i)\,,
\end{cases}\\
\label{ad2}
& \widehat{R}_{_{ij}}=
\begin{cases}
\widehat{\delta}_{_{ij}}, & j \in N_{J}^{I}(i) \bigcap N_{J}^{O}(i)\\
\frac{\zeta+J+1}{2}, & j \notin N_{J}^{I}(i) \bigcap N_{J}^{O}(i)\,,
\end{cases}
\end{align}
where 
\begin{align*}
& \delta_{_{ij}}=\#\left\{k\in N^{I}_{_J}(i) \bigcap N^{O}_{_J}(i):\,\left\|\mathbf{x}_{i}-\mathbf{x}_{k}\right\|\leq\left\|\mathbf{x}_{i}-\mathbf{x}_{j}\right\|\right\}\\
& \widehat{\delta}_{_{ij}}=\# \left\{k\in N^{I}_{_J}(i) \bigcap N^{O}_{_J}(i):\,\left\|\widehat{\mathbf{y}}_{i}-\widehat{\mathbf{y}}_{k}\right\|\leq\left\|\widehat{\mathbf{y}}_{i}-\widehat{\mathbf{y}}_{j}\right\|\right\}\\
& \zeta=\# \left(N_{J}^{I}(i) \bigcap N_{J}^{O}(i) \right).
\end{align*} 
The trimming in \eqref{ad1} and \eqref{ad2} is to make the ranks comparisons local. To measure the output error, we can define the following. 

\theoremstyle{definition} 
\newtheorem{defi}{Definition}
\begin{defi} {\it Local rank correlation for output error }:

For a given point $\mathbf{x}_i$, define the local output Spearman correlation as
\begin{align}
\label{rhoO}
\rho^{O}_{_J}(i,\mathbf{X},\widehat{\mathbf{Y}})=1-\frac{6\left( \underset{j \in N_{_J}^{O}(i)}{ \sum}\left\{\left(S_{_{ij}}-\widehat{r}_{_{ij}}\right)^{2}\right\}+U\right)}{J(J^{2}-1)}\,,
\end{align}
where $U=\left[(J-\zeta)^{3}-(J-\zeta)\right]/12$ is the adjustment made for the appearance of ties \citep{kendall1}. Define the local output Kendall correlation as
\begin{align}
\label{tauO}
\tau^{O}_{_J}(i,\mathbf{X},\mathbf{Y})=\frac{\underset{j<k \in N_{_J}^{O}(i)}{\sum}2\,\mathrm{sign}\left\{(S_{_{ij}}-S_{_{ik}})\cdot(\widehat{r}_{_{ij}}-\widehat{r}_{_{ik}})\right\}}{J(J-1)}.
\end{align}

For a given input dataset $\mathbf{X}$ and a given dimensionality reduction method $\widehat{\psi}: \mathbf{X} \mapsto \widehat{\psi}(\mathbf{X})$, an overall goodness measure can be defined by averaging the local correlation over all cases in the sample. 
\begin{align}
\label{GO}
G^{O}_{_J}(\widehat{\psi},\mathbf{X})=\frac{1}{n}\sum_{i=1}^{n}\Gamma^{O}_{_J}\left( i,\mathbf{X},\widehat{\psi}(\mathbf{X}) \right),
\end{align}
where $\Gamma^{O}_{_J}$ can be either $\rho^{O}_{_J}$, or $\tau^{O}_{_J}$.
\end{defi}

The local rank correlations $\rho^{O}_{_J}(i)$ or $\tau^{O}_{_J}(i)$ measure the similarity, in terms of output errors, between the corresponding neighborhoods, $N_{_J}^{I}(i)$ and $N_{_J}^{O}(i)$. Similarly, we can define local rank correlations to measure the input error.

\begin{defi} {\it Local rank correlation for input error} :

Given an input dataset $\mathbf{X}$ and a low-dimensional representation $\widehat{\mathbf{Y}}$, the local Spearman correlation and local Kendall correaltion for the input error at the $i$-th case are defined as
\begin{align}
& \rho^{I}_{_J}(i,\mathbf{X},\widehat{\mathbf{Y}})=1-\frac{\underset{j \in N_{_J}^{I}(i)}{\sum}\left\{\left(s_{_{ij}}-\widehat{R}_{_{ij}}\right)^{2}\right\}+U}{\frac{1}{6}(J^{3}-J)}\label{rhoI}\,,\\
& \tau^{I}_{_J}(i,\mathbf{X},\widehat{\mathbf{Y}})=\frac{\underset{j<k,\, j,\,k \in N_{_J}^{I}(i)}{\sum}\mathrm{sign}\left\{(\widehat{R}_{_{ij}}-\widehat{R}_{_{ik}})\cdot(s_{_{ij}}-s_{_{ik}})\right\}}{\frac{1}{2}J(J-1)}\label{tauI}\,.
\end{align}

The overall goodness measure of a given method $\widehat{\psi}$ and input data $\mathbf{X}$ is defined as
\begin{align}
\label{GI}
G^{I}_{_J}(\widehat{\psi},\mathbf{X})=\frac{1}{n}\sum_{i=1}^{n}\Gamma^{I}_{_J}\left( i,\mathbf{X},\widehat{\psi}(\mathbf{X}) \right)\,,
\end{align}
where $\Gamma^{I}_{_J}$ can be either $\rho^{I}_{_J}$, or $\tau^{I}_{_J}$. 
\end{defi}
\subsection{Remark}

The proposed local rank correlations have some nice properties. The higher values of local measures $\Gamma^{I}_{_J}(i)$ and $\Gamma^{O}_{_J}(i)$ indicate a higher degree of similarity between the original data and the low-dimensional configuration in the neighborhood of case $i$, while values close to 0, or negative values indicate that low-dimensional configuration fails to preserve the local structure of the input data in certain neighborhoods. Two special situations are:
\begin{itemize}
\item $\Gamma^{I}_{_J}(i)=\Gamma^{O}_{_J}(i)=1$ if all the ranking relationships of the observed data $\mathbf{X}$ in the neighborhood of case $i$ are preserved exactly in the corresponding neighborhood in the output data $\widehat{\mathbf{Y}}$.

\item The expected values $\mathrm{E}\left[\,\Gamma^{I}_{_J}(i)\,\right]$ and $\mathrm{E}\left[\,\Gamma^{O}_{_J}(i)\,\right]$ are both zero, for any case $i$, where the output $\widehat{\mathbf{Y}}$ is generated by an algorithm which is stochastically independent of the input data $\mathbf{X}$.
\end{itemize} 
These two facts hold for both local Spearman and Kendall correlations. 
Notice that the second situation is worse than we can have in practice. Moreover,  the local measures $\Gamma^{I}_{_J}(i)$ and $\Gamma^{O}_{_J}(i)$, can achieve negative values for some $i$. Nevertheless, the overall goodness measures $G_{_J}^{O}$ and $G_{_J}^{I}$, for sensible algorithms, will take values between 0 and 1. We remind the reader that the use of ranks is to protect against non-normality and extreme skewness of distance distributions in high dimensions. 

The computational complexity is also of interest. To calculate the goodness measure, we first construct the $J$-nearest neighbor graph for both $\mathbf{X}$ and $\widehat{\mathbf{Y}}$. This step scales as $O(n^{2}p)$. In the next step, we calculate the local rank correlation in each neighborhood. This scales (in each neighborhood) as $O(J)$ for Spearman $\rho_{_J}$ and $O(J\log{J})$ for Kendall $\tau_{_J}$. Therefore, since $J\leq n$, the total complexity of calculating $G^{I}_{_J}$ (or $G^{O}_{_J}$) for $\rho_{_J}$ scales as $O(n^{2}p)$. The total complexity of calculating $G^{I}_{_J}$ (or $G^{O}_{_J}$) for $\tau_{_J}$ scales as  $O(n^{2}p+nJ\log J)$. 

To use the proposed goodness measure $G_{_J}$ for assessing the performance of a dimension reduction method, four local measures can be chosen. We may choose either $\Gamma^{I}_{_J}$ or $\Gamma^{O}_{_J}$, and we may also  choose to use either Spearman $\rho_{_J}$ or Kendall $\tau_{_J}$. The measures $\Gamma^{I}_{_J}$ and $\Gamma^{O}_{_J}$ quantify different types of errors in dimension reduction. Although these two types of errors usually occur together, having both $G^{I}_{_J}$ and $G^{O}_{_J}$ provides more complete information about the performance of a given method. 

\subsection{Choice of \texorpdfstring{$J$}{J}}

In the proposed measures, $J$ is a user-specified tuning parameter, which specifies the neighborhood size for local rank comparisons. Notice that some nonlinear dimensionality reduction methods start with a $K$-nearest neighbors graph, and $K$ is also a user-specified parameter. The choice of $J$ in the local rank correlation does not have to depend on the value of $K$. Ideally, $J$ is chosen sufficiently small that there exists a $q$-flat $F^q$ such that the data points in $N_{_J}^I(i)$ lie approximately on $F^q$. In addition, $J$ is ideally chosen sufficiently large that each $N_{_J}^I(i)$ is informative about the local geometrical characteristics of the data. In practice, we expect $G_{_J}(\widehat{\psi},\mathbf{X})$ to be large and stable for values of $J$ sufficiently small to satisfy the former condition. So we can choose $J$ by plotting $G_{_J}(\widehat{\psi},\mathbf{X})$ against $J$ and choosing the largest value of $J$ for which the stability is observed. 

\subsection{Adjustments for output-normalized methods}

Normalization of the output, as used in algorithms such as Local Linear Embedding \citep{lle}, Laplacian Eigenmap \citep{LEM}, and Local Tangent Space Alignment \citep{ltsa}, affinely transforms the geometrical structure of neighborhoods, so that the distance relationships will not be preserved between the input and the output configurations. It is not adequate to check the correspondence of nearest neighbor ranks between $\mathbf{X}$ and $\widehat{\psi}(\mathbf{X})$ for those output-normalized methods. Instead, we will look for a transformation matrix $\widehat{\mathbf{A}}_{q\times q}$, and assess the performances of output-normalized methods by an adjusted measure
\begin{align}
\label{adj}
G^{\mathbf{A}}_{_J}(\widehat{\psi},\mathbf{X})=\frac{1}{n}\sum_{i=1}^{n}\Gamma_{_J}(i,\mathbf{X},\widehat{\psi}_{_\mathbf{A}}(\mathbf{X}))\,,
\end{align}
where $\widehat{\psi}_{_\mathbf{A}}(\mathbf{X})=\widehat{\psi}(\mathbf{X})\cdot \widehat{\mathbf{A}}$, and $\Gamma_{_J}$ can be $\rho^{I}_{_J}$, $\rho^{O}_{_J}$ or $\tau^{I}_{_J}$, $\tau^{O}_{_J}$.

It is hoped that after the affine transformation $\widehat{\mathbf{A}}$, $\widehat{\psi}_{_\mathbf{A}}(\mathbf{X})$ can preserve the proximities between neighboring points as much as possible, i.e. $\widehat{\mathbf{A}}$ will minimize the least squared error
$$\sum_{i=1}^{n}\sum_{j \in N^{^I}_{_J}(i)}\left[ (\mathbf{x}_{i}-\mathbf{x}_{j})^{\prime}(\mathbf{x}_{i}-\mathbf{x}_{j})-(\mathbf{y}_{i}-\mathbf{y}_{j})^{\prime}\mathbf{A}^{\prime}\mathbf{A}(\mathbf{y}_{i}-\mathbf{y}_{j}) \right]^{2}\,,$$
where $\mathbf{x}_{i}$ and $\mathbf{y}_{i}$ are the corresponding points in the original data and in the output of the algorithm $\widehat{\psi}$, respectively.

\section{Experiments}
\label{sec4}

In this section, we conduct numerical examples on some benchmark datasets to illustrate the effectiveness of the local rank correlation. 

\theoremstyle{definition} 
\newtheorem{example}{Example}
\begin{example} {\it The Swiss roll and the S-curve:} 
In this experiment, $n=1000$ data points are generated randomly from two manifolds, the Swiss roll and the S-curve. They are both 2-dimensional manifolds embedded into $\mathbb{R}^{3}$ (Figure \ref{gen}). The data
points are colored to help readers recognize the structure of the manifolds. Among many dimension reduction methods, we choose four diverse methods, namely   ISOMAP, LTSA, MVU, and PCA. Figure \ref{swiss} and Figure \ref{Scurve} show four output configurations in $\mathbb{R}^2$ from these methods for the Swiss roll and the S-curve, respectively. We evaluate the performance of four methods by the local rank correlations $G^{I}_{_J}(\widehat{\psi},\mathbf{X})$ and $G^{O}_{_J}(\widehat{\psi},\mathbf{X})$ with both Spearman $\rho_{_J}$ and Kendall $\tau_{_J}$. Notice that LTSA is an output-normalized method, and therefore, its performance is assessed by the adjusted measures \eqref{adj}. 

The goodness measures are calculated under different values of $J$. Figure \ref{swiss-lcorr} and Figure \ref{scurve-lcorr} show the values of $G_{_J}(\widehat{\psi},\mathbf{X})$ for each method as functions of $J$.

\begin{figure}[!ht]
	\centering
		\includegraphics[width=0.8\textwidth]{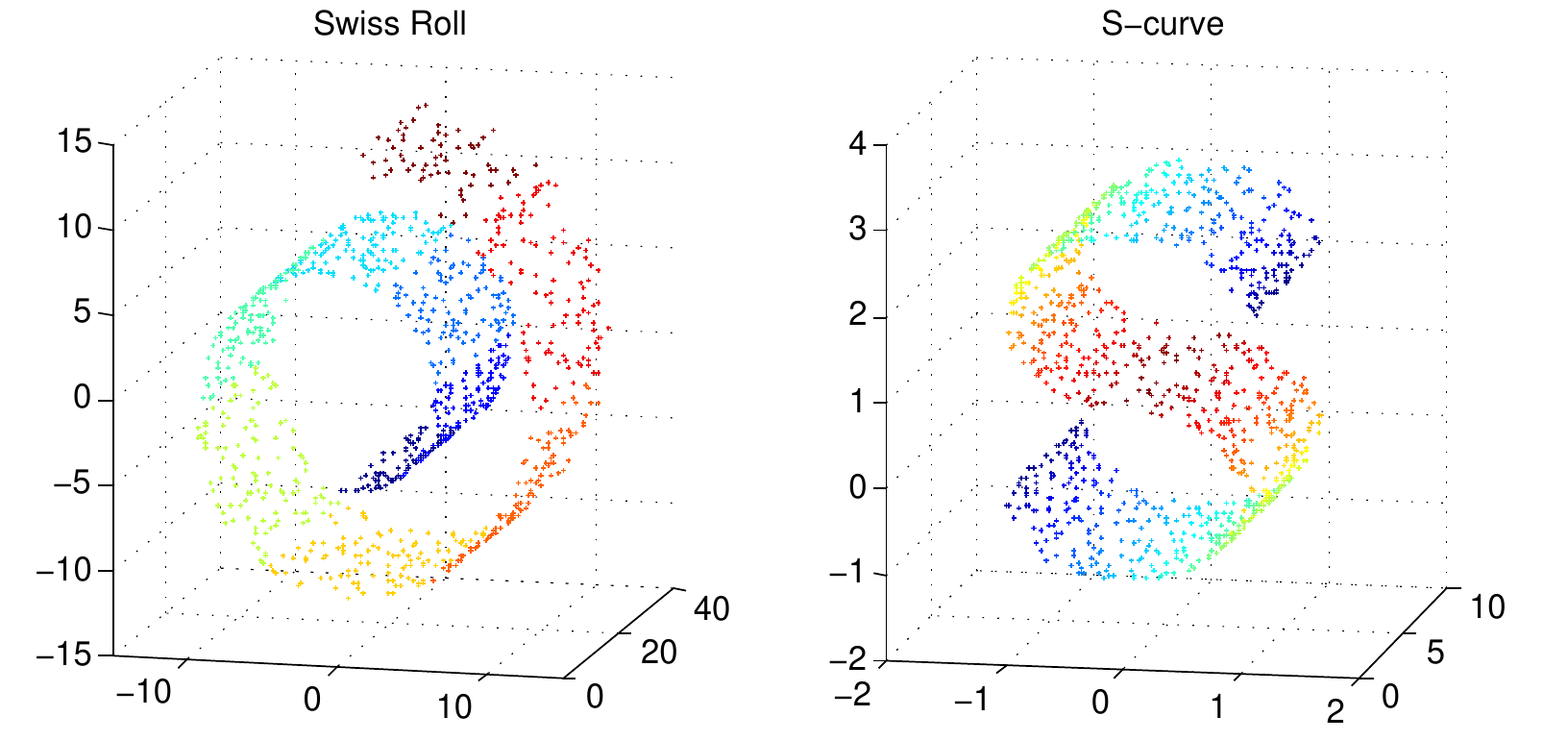}
	\caption{Data on the Swiss roll and the S-curve.}
	\label{gen}
\end{figure}

\begin{figure}[!ht]
	\centering
		\includegraphics[width=0.8\textwidth]{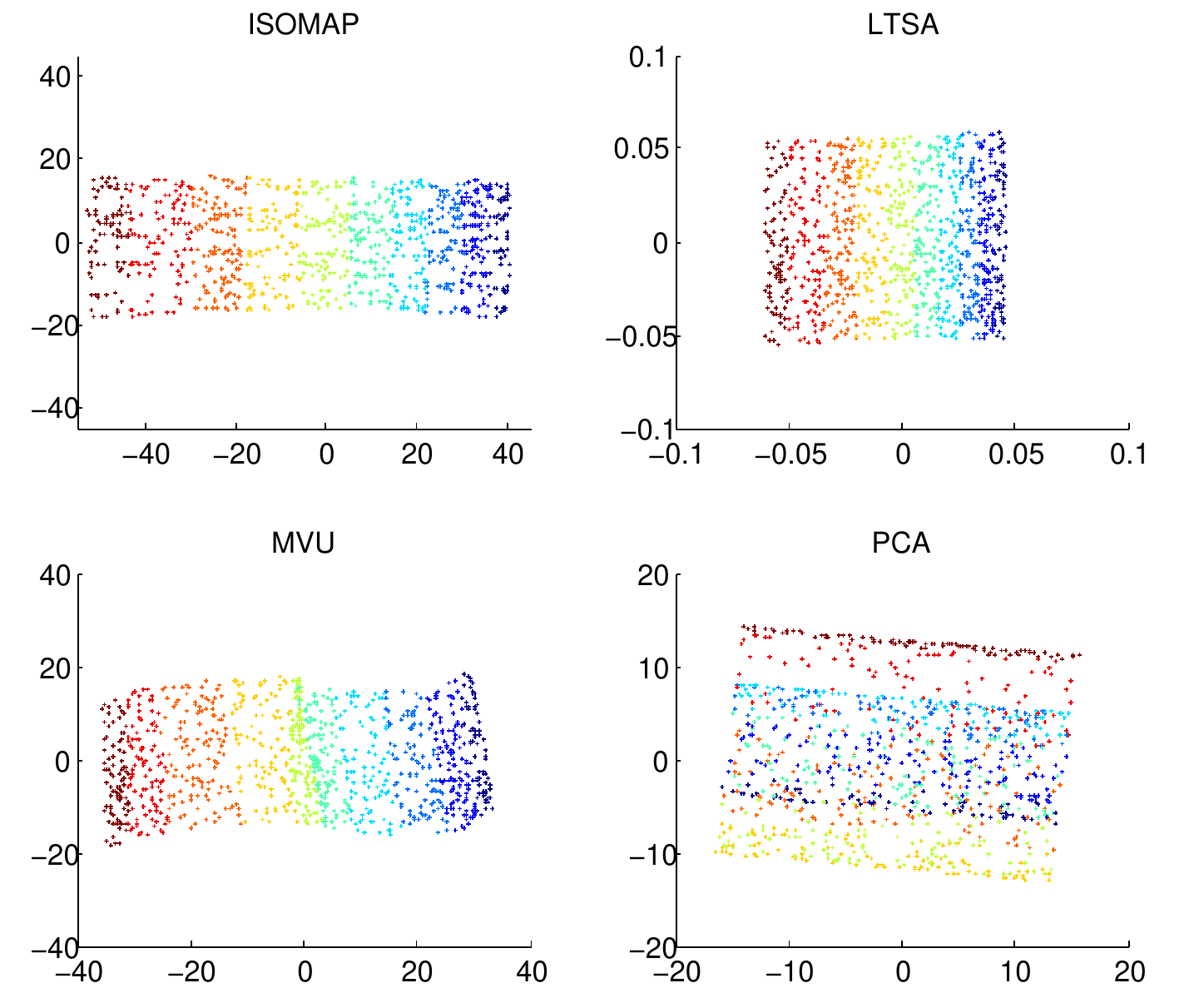}
	\caption{Two-dimensional output configurations of different methods for the Swiss roll.}
	\label{swiss}
\end{figure}

\begin{figure}[!ht]
	\centering
		\includegraphics[width=0.8\textwidth]{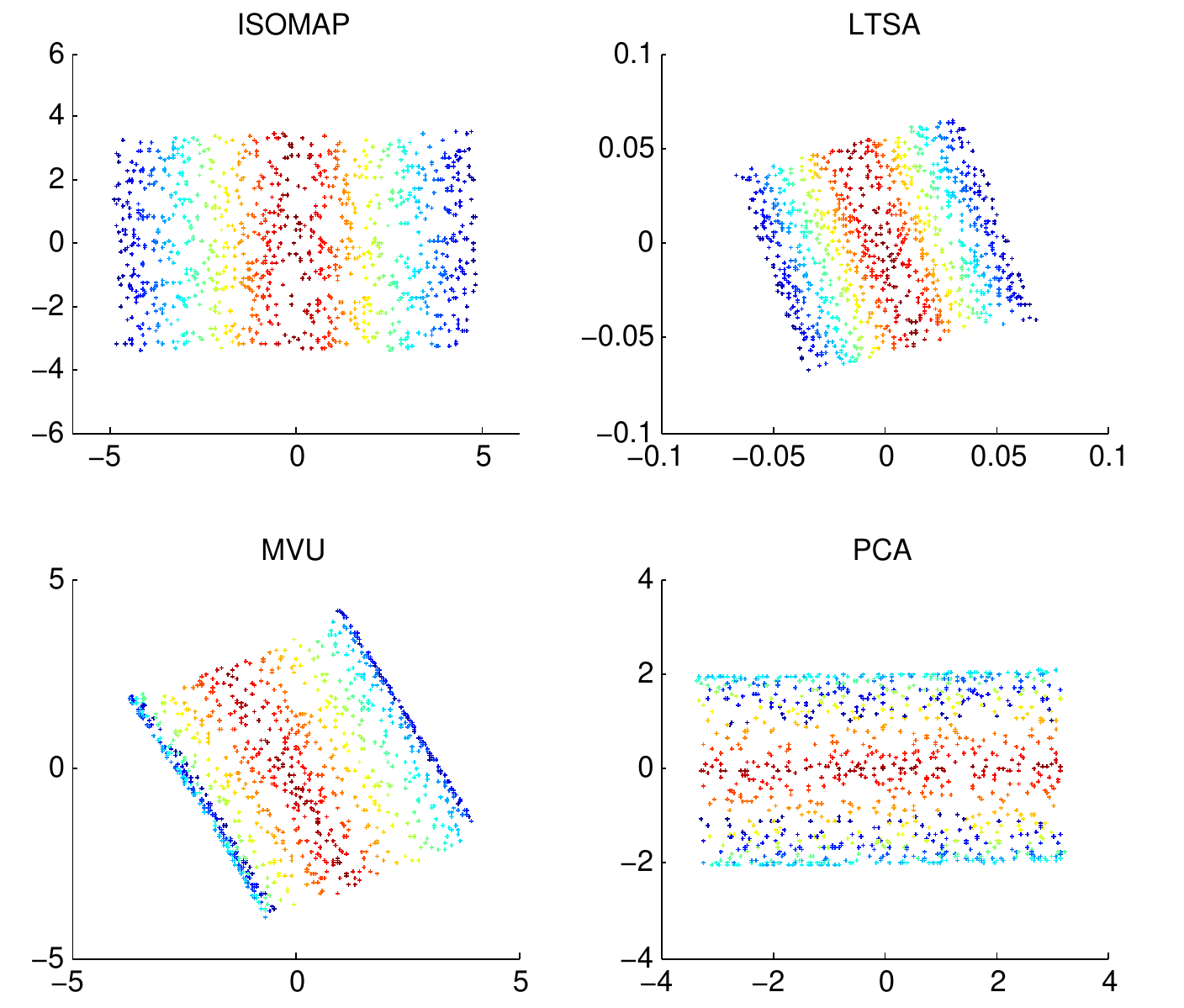}
	\caption{Low-dimensional configurations of different methods for S-curve.}
	\label{Scurve}
\end{figure}

\begin{figure}[!ht]
	\centering
		\includegraphics[width=1\textwidth]{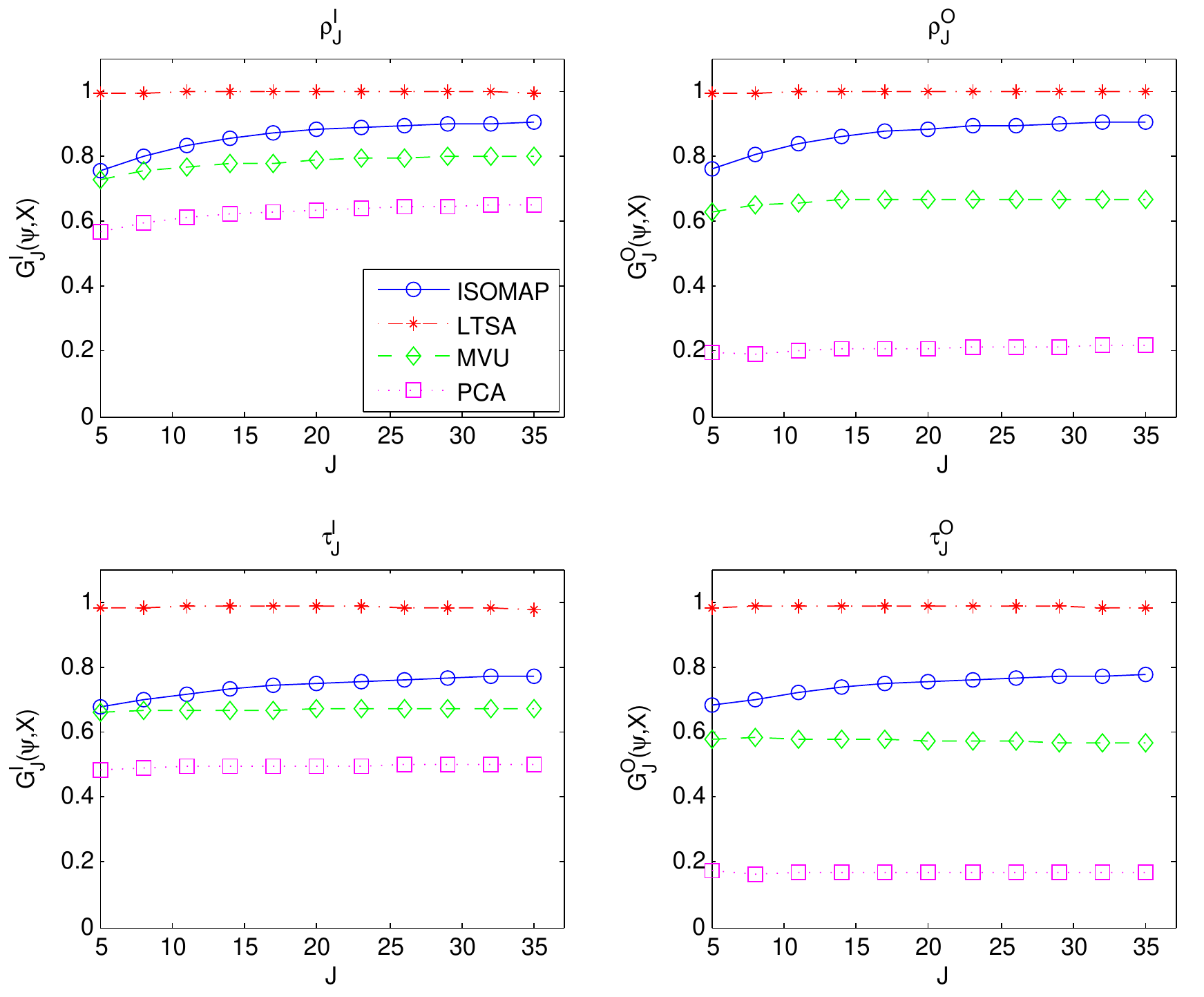}
	\caption{Local Spearman correlation as functions of $J$.}
	\label{swiss-lcorr}
\end{figure}

\begin{figure}[!ht]
	\centering
		\includegraphics[width=1\textwidth]{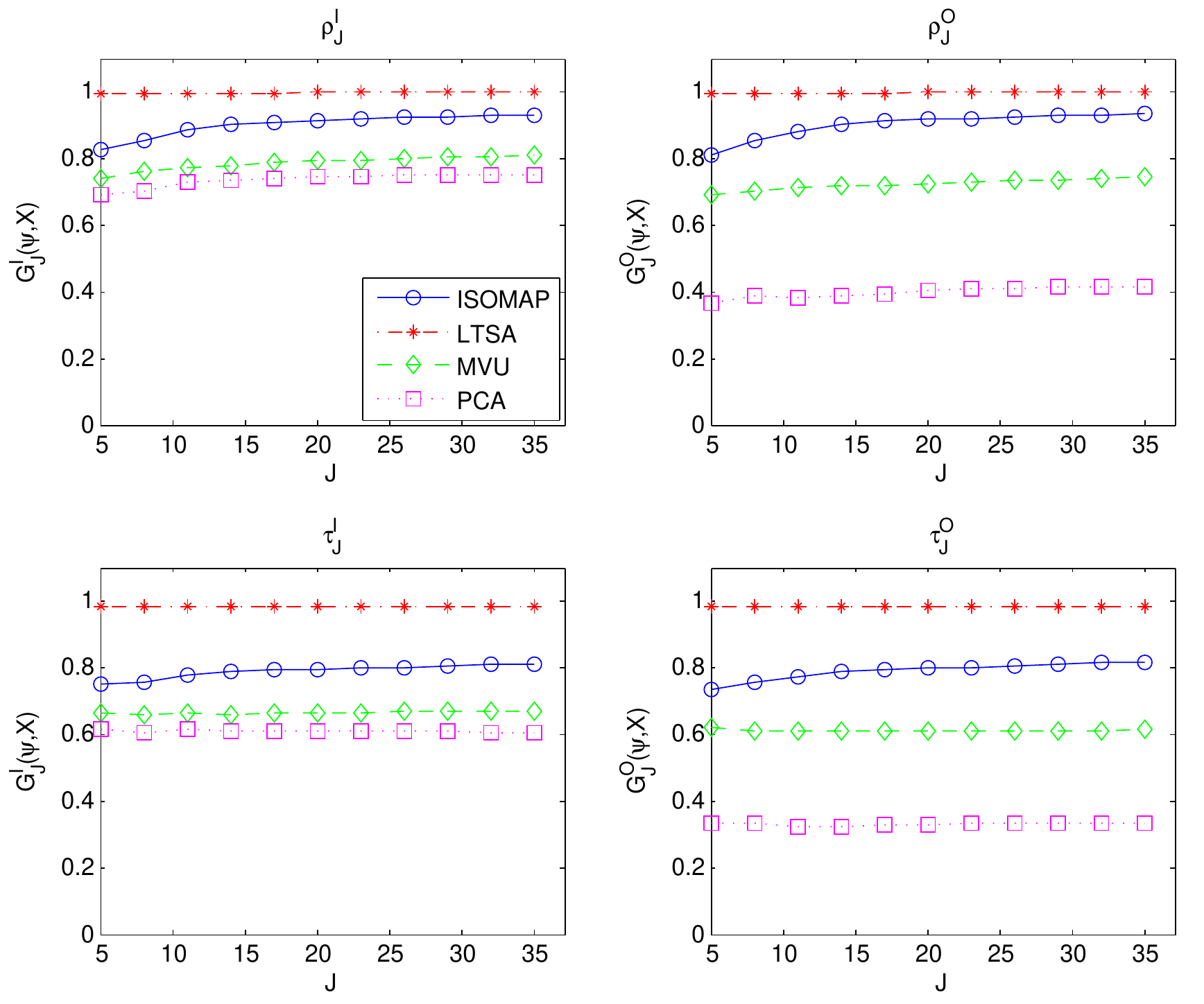}
	\caption{Local Spearman correlation as functions of $J$.}
	\label{scurve-lcorr}
\end{figure}

As can be seen from Figure \ref{swiss} and Figure \ref{Scurve}, in the 2-dimensional configurations from PCA in both the Swiss roll and the S-curve, the points with different colors are mixed together, because PCA fails to recover the nonlinear structure of the embedded data. Among three nonlinear methods, the configurations from MVU preserve the structure to some extent. Points with different colors are reasonably separated in the middle, but they mix a little at boundaries. Both LTSA and ISOMAP preserve the color level well, indicating a better embedding than MVU and PCA. These facts are all correctly reflected by the four goodness measures $G_{_J}(\widehat{\psi},\mathbf{X})$ in Figure \ref{swiss-lcorr} and Figure \ref{scurve-lcorr}. Also, both figures show that all the four measures are stable within a reasonable range of $J$.

We also compare the local rank correlation (LRC) with the goodness measures, MRREs, T\&C, and LCMC (all with $J=6$). The results are reported in Table \ref{tab-swiss} and Table \ref{tab-Scurve}. 
\begin{table}[!ht]
\large
    \begin{center}
        \begin{tabular}{|c|c|c|c|c|c|c|c|c|c|}
        \hline
\multirow{2}{*}{Methods} & \multicolumn{4}{|c|}{LRC} & \multicolumn{2}{|c|}{MRREs} & \multicolumn{2}{|c|}{T \& C} & \multirow{2}{*}{LCMC}\\ \cline{2-9}
& $\rho^{I}_{_J}$ & $\rho^{O}_{_J}$ & $\tau^{I}_{_J}$ & $\tau^{O}_{_J}$ & $M_{_O}$ & $M_{_I}$ & $T$ & $C$ & \\
            \hline \hline
ISOMAP & 0.787 & 0.782 & 0.701 & 0.698 & 0.999  &  0.999  &  0.999  &  0.999  & 0.894\\ 
LTSA & 0.988 & 0.978 & 0.981 & 0.975 & 0.999  &  0.998 &   0.993  &  0.998  &  0.609\\
MVU &  0.703 & 0.623 & 0.653 & 0.578 & 0.999  &  0.999  &  0.996  &  0.999  &  0.828\\
PCA &  0.594 & 0.198 & 0.483 & 0.171 & 0.998  &  0.997  &  0.883  &  0.995  &  0.415\\              \hline \hline          
        \end{tabular}
    \end{center}
	\caption{Assessing ISOMAP, LTSA, MVU, PCA in Swiss Roll data ($J=6$).}
	\label{tab-swiss}
\end{table}
\begin{table}[!ht]
\large
    \begin{center}
        \begin{tabular}{|c|c|c|c|c|c|c|c|c|c|}
        \hline
\multirow{2}{*}{Methods} & \multicolumn{4}{|c|}{LRC} & \multicolumn{2}{|c|}{MRREs} & \multicolumn{2}{|c|}{T \& C} & \multirow{2}{*}{LCMC}\\ \cline{2-9}
& $\rho^{I}_{_J}$ & $\rho^{O}_{_J}$ & $\tau^{I}_{_J}$ & $\tau^{O}_{_J}$ & $M_{_O}$ & $M_{_I}$ & $T$ & $C$ & \\
            \hline \hline
ISOMAP & 0.816 & 0.804 & 0.763 & 0.803 & 0.999 & 0.999  &  1.000  &  1.000  &  0.891\\ 
LTSA & 0.994 & 0.993 & 0.983 & 0.979 & 0.999  & 0.999 &  0.999  &  0.999  &  0.867\\
MVU & 0.721 & 0.646 & 0.695 & 0.617 & 0.999  & 0.998 &  0.993  &  0.997  &  0.754\\
PCA & 0.673 & 0.375 & 0.388 & 0.369 & 0.998  &  0.998  &  0.963   & 0.998  &  0.584\\              \hline \hline          
        \end{tabular}
    \end{center}
	\caption{Assessing ISOMAP, LTSA, MVU, PCA in S-curve data ($J=6$).}
	\label{tab-Scurve}
\end{table}
%%%%%%%%%%%%

We now turn to the analysis of the Swiss roll data.  We examine the output visually first.  As expected, PCA works poorly. This is a consequence of the many-to-one nature of a linear projection of the Swiss roll.  All three of the other algorithms separate colors well.  However, an ideal output should be perfectly rectangular or square.  Visually, we prefer LTSA to ISOMAP and prefer ISOMAP to MVU.  

We next turn to the performance measures for the Swiss roll output. The LCMC performance measure has identified the failure of PCA to account for the nonlinearity of the Swiss roll.  However, it also ranks the outcome of MVU as superior to LTSA, which is not visually supported in Figure \ref{swiss}.  The ISOMAP algorithm is ranked highest by LCMC, but only slightly higher than MVU.  Despite the fact that the output from LTSA and ISOMAP are visually close, the LCMC measures are quite distinct.  It would seem reasonable to assume that if LCMC is picking up problems with these algorithms, the problems are not visually obvious.

The trustworthiness and continuity measures, present a more complex picture.  As expected, all algorithms perform reasonably well on the continuity criterion.  For example, the linear projection defining the outcome of PCA is a continuous mapping of the Swiss roll, and therefore satisfies the continuity criterion well with a high value of C.  That is also the case for the other algorithms.  The trustworthiness measure clearly separates out PCA as the least desirable algorithm, as expected.  Other algorithms perform very well and similarly to each other.  Once again, MVU is ranked higher than LTSA, in contrast to our visual interpretation.

Turning to MRREs, we see that very little separation can be seen among the algorithms.  Since these measures are not standardized, we must be wary of drawing too many conclusions from the proximity of these values to one.  Nevertheless, MVU is again ranked higher than LTSA and is not separated in performance from ISOMAP.

We are happy with the correspondence between LRC and our conclusions based upon visual inspection.  The LTSA algorithm is ranked highest by all four measures of LRC, followed by ISOMAP, MVU and PCA.  We find it especially helpful that there is no inconsistency in ranking between the four measures of LRC.

Our conclusions, both visual and quantitative, for the S-curve trend in a similar direction to the Swiss roll.  Visually, we rank ISOMAP and LTSA best with little to choose between them.  The MVU algorithm works very well, but performs poorly close to the boundary, where a mixing of colors can be seen.  The PCA algorithm is, once again, the worst of the four, but within an acceptable standard as the output is not from a many-to-one projection. 

Next, we examine the performance measures for the S-curve output.  Note that LCMC performs in close correspondence with our visual analysis.  Similar remarks hold for T \& C.  Although continuity ranks PCA better than MVU, we should note that PCA is a continuous algorithm and that this is reasonable.  Trustworthiness rankings correspond to the visual rankings.  The discrimination provided by MRREs is, again, unclear.  Nevertheless it is consistent with visual conclusions.  It is reasonable to conclude from the Swiss roll data and S-curve data that MRREs are less useful for these two benchmark examples.  Under the LRC measures, LTSA is consistently the best, followed by ISOMAP, MVU and PCA in that order.  This agrees with the visual analysis.
\end{example}

\section{Choosing tuning parameters for algorithms}

Many nonlinear dimension reduction methods start with constructing the $K$-nearest neighbor graph. The success of graph-based nonlinear dimensionality reduction methods depends heavily on the selection of $K$.  
If $K$ is chosen to be too small, the local geometric structure cannot be accurately represented in the neighborhood graph. On the other hand, if $K$ is chosen to be too large, the $K$-nearest neighbor graph will
contain \textit{shortcuts}, i.e. two points will be mistakenly considered as neighbors when they are in fact far away on the manifold. In practice, $K$ is usually chosen by experience or trial and error.

The local rank correlation can help in choosing $K$, because $G_{_J}(\widehat{\psi},\mathbf{X})$ measures the performance of $\widehat{\psi}$. We may apply the algorithm over a range of values of $K$, and calculate $G_{_J}(\widehat{\psi},\mathbf{X})$ as a function of $K$ (as shown in Figure \ref{fig51}). We then pick the $K$ that corresponds to the largest $G_{_J}(\widehat{\psi},\mathbf{X})$.

\begin{example}{\it Selecting neighborhood size $K$ in ISOMAP:} 
Here we consider the performance of ISOMAP on the Swiss roll manifold. We demonstrate that it is risky to make a desultory choice of $K$, and how local rank correlation can solve this problem. The data are generated randomly on the Swiss roll manifold with sample size $n=1500$. The ISOMAP algorithm is applied on the data with different values of $K$, and Figure \ref{fig52} shows the respective low-dimensional configurations. In each case, the performance is evaluated by the local rank correlation and displayed as a function of $K$ in Figure \ref{fig51}.

\begin{figure}[!ht]
	\centering
		\includegraphics[width=0.8\textwidth]{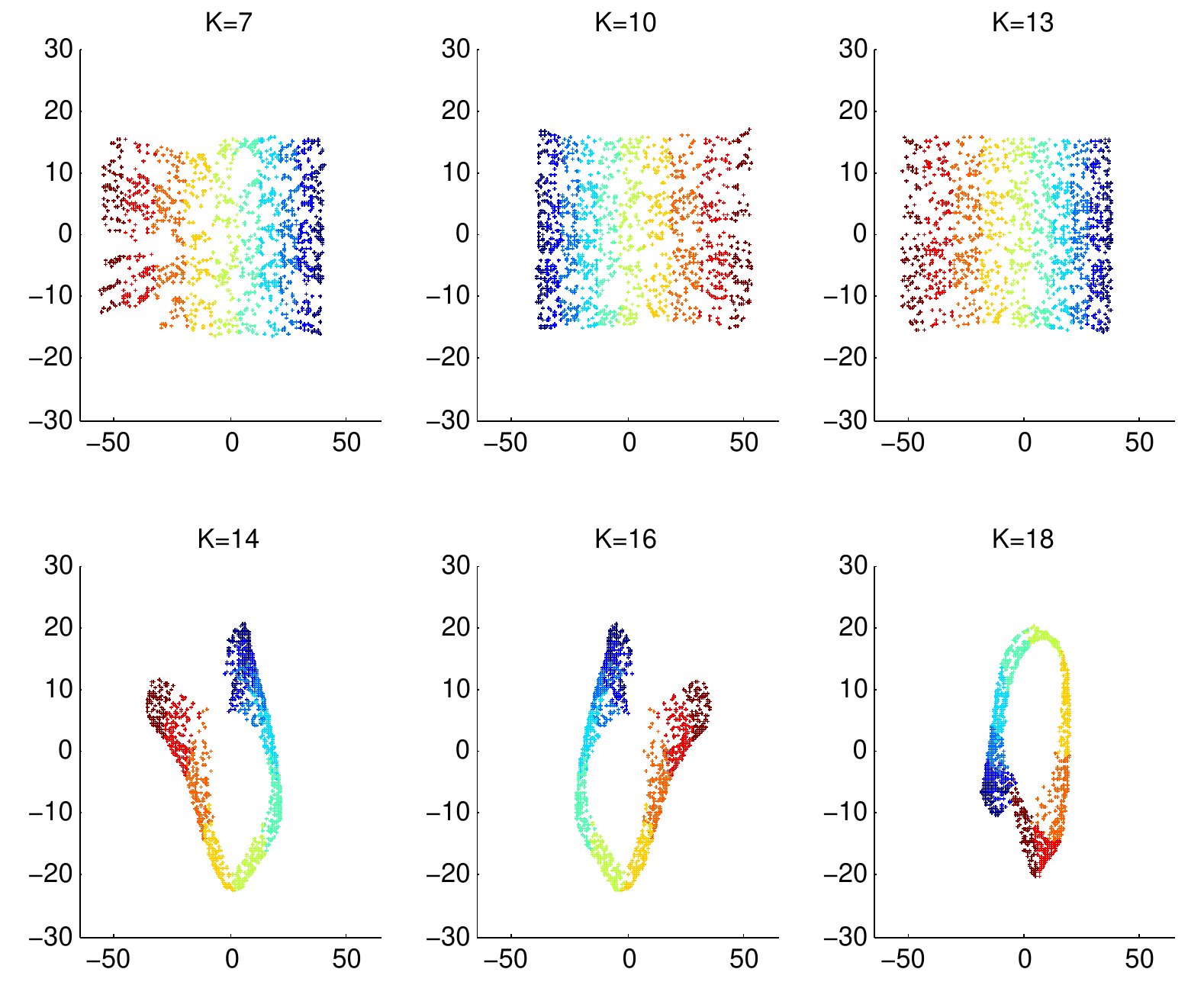}
	\caption{Low-dimensional configurations with different values of $K$.}
	\label{fig52}
\end{figure}

\begin{figure}[!ht]
	\centering
		\includegraphics[width=1\textwidth]{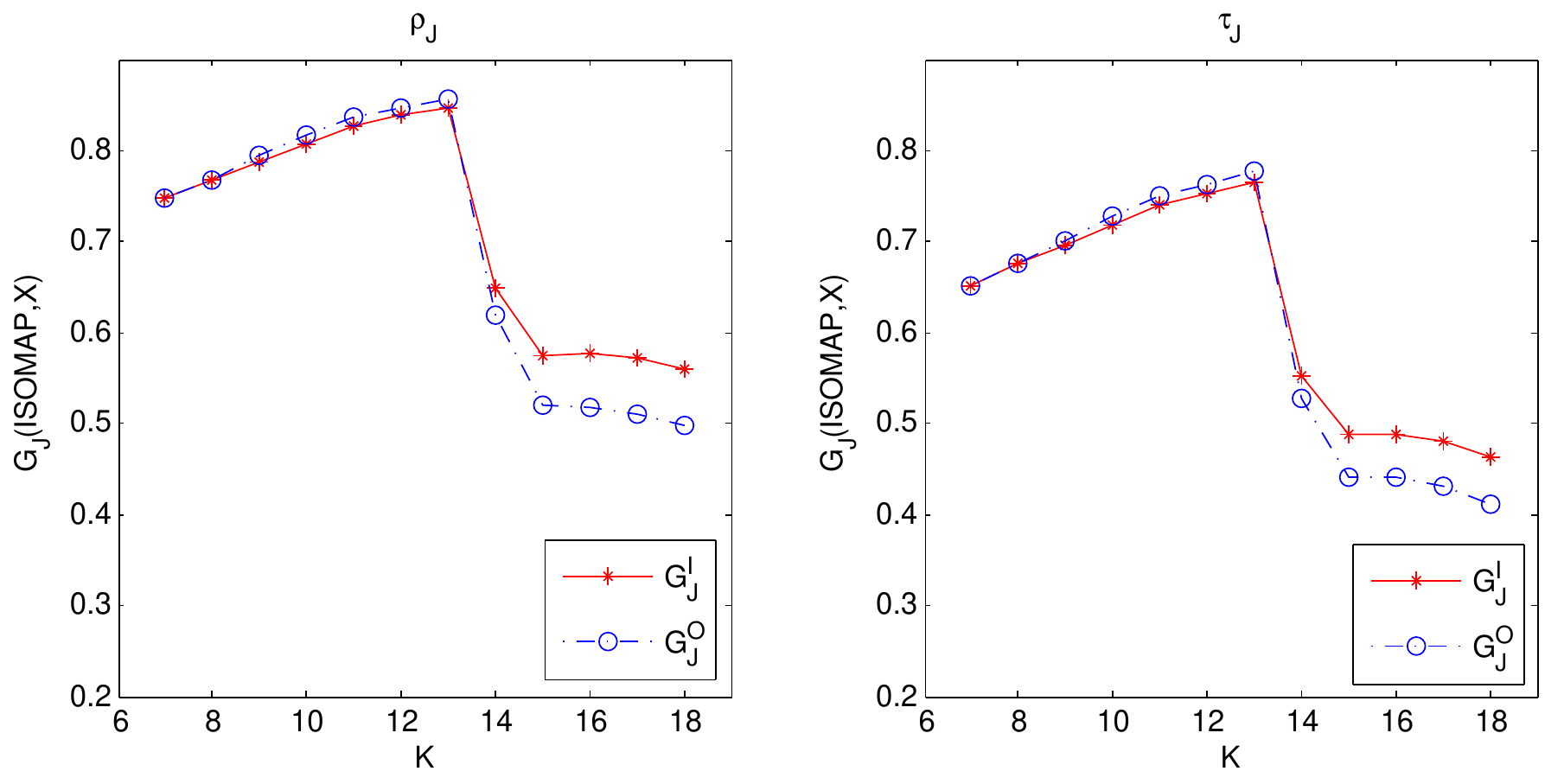}
	\caption{Local rank correlation as a function of $K$ ($J=6$).}
	\label{fig51}
\end{figure}

In Figure \ref{fig51}, the left panel shows $G^{I}_{_J}$ and $G^{O}_{_J}$ with Spearman $\rho_{_J}$, and the right panel shows $G^{I}_{_J}$ and $G^{O}_{_J}$ with Kendall $\tau_{_J}$. 
As can be easily noticed in Figure \ref{fig52}, the performance of ISOMAP gets better as $K$ increases from $K=7$ to $K=13$. A crucial change has happened at points $K=13$ and $K=14$. In these two situations, the neighborhood sizes only differ by 1 but the corresponding configurations suddenly become unsatisfactory (at $K=14$). 
The fact is correctly captured by the local rank correlation and reflected in Figure \ref{fig51}. In all four measures, we observe a peak at $K=13$, and a steep drop at $K=14$. For nonlinear methods which contain the neighborhood size $K$ as a tuning parameter, it is often desirable to choose a relatively large value of $K$ to get a better embedding. On the other hand, a value of $K$ which is too large will invalidate the procedure.

\end{example}

\section{Estimating the intrinsic dimensionality of a manifold}

Another key parameter in dimension reduction algorithms is the intrinsic dimensionality $q$. The local rank correlation can be applied to help in estimating the intrinsic dimensionality. The idea is that if the dimensionality of the low-dimensional representation is chosen to be too small, important features of the original data might be ``collapsed'' onto the same dimensions so that the topological structure cannot be preserved very well. As the dimensionality of the representation increases, while remaining below the correct dimension $q$ of the manifold, the local rank correlation should increase. On the other hand, when the dimensionality of the representation is greater than $q$, the only additional information in the data provided by the additional dimensions would be noise. Therefore, provided that the noise is small, the local rank correlation would become stable at values larger than $q$. This is similar to the scree plot (\cite{ScreePlot}), but in reverse, used for choosing dimensionality of linear manifolds in PCA.   
In practice, for a given dataset $\mathbf{X}$ and a chosen method $\widehat{\psi}$, we may apply the method with different dimensions, and evaluate the performance of $\widehat{\psi}$ by $G_{_J}(\widehat{\psi},\mathbf{X})$. We estimate the intrinsic dimensionality by plotting $G_{_J}(\widehat{\psi},\mathbf{X})$ as a function of dimension, and choosing the value $q$, beyond which $G_{_J}(\widehat{\psi},\mathbf{X})$ becomes stable.

\begin{example}{\it Estimating the intrinsic dimensionality of the sculpture face data}: 
The sculpture face dataset (\cite{isomap}) includes 698 images, each image having $64\times 64$ pixels of a sculpture face while varying three free parameters: left-right pose, up-down pose, and lighting direction. So the data are originally in $\mathbb{R}^{64\times 64}$.  The sculpture face images are recorded as $64 \times 64$ vectors. Since images are taken from the same sculpture face by varying three parameters, the intrinsic dimensionality of the manifold on which these data vectors lie is three. We apply the ISOMAP algorithm with different values of $q$ having chosen the neighborhood size $K=8$. The local rank correlations are calculated as functions of $q$. In Figure \ref{fig54}, the left panel shows $G^{I}_{_J}$ and $G^{O}_{_J}$ with Spearman $\rho_{_J}$, and the right panel shows $G^{I}_{_J}$ and $G^{O}_{_J}$ with Kendall $\tau_{_J}$. As can be seen, all four curves become stable beyond $q=3$, based on which we estimate the intrinsic dimensionality to be $\widehat{q}=3$. 

\begin{figure}[!ht]
	\centering
		\includegraphics[width=1\textwidth]{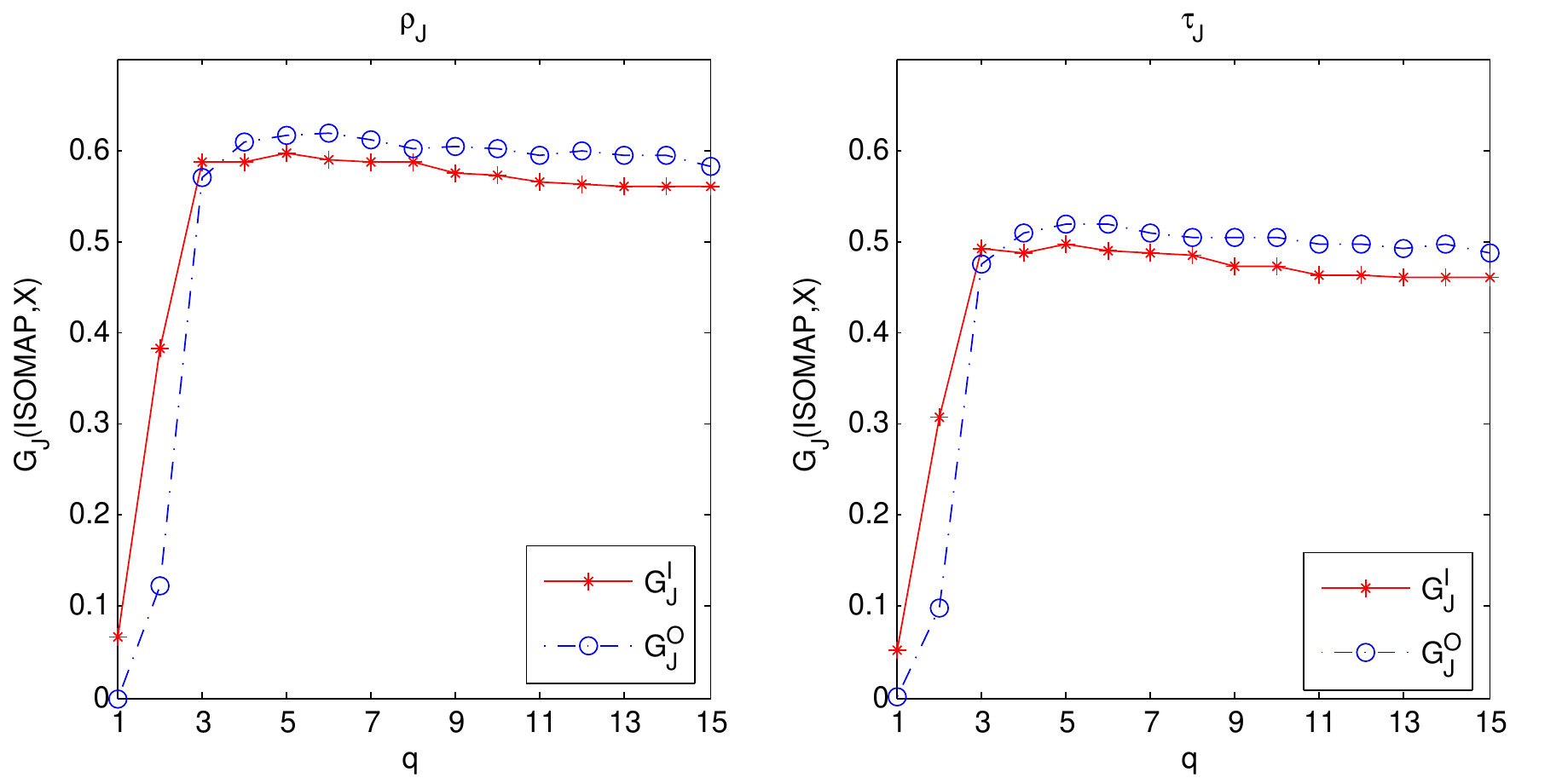}
	\caption{Local rank correlation as a function of dimensionality $q$ ($J=6$).}
	\label{fig54}
\end{figure}

\end{example}

\section{Conclusions} 
To quantify the performance of a dimension reduction method, we introduced a family of local rank correlation measures, which are   
easily interpretable and motivated by their robustness properties for nearest neighbor distributions in high dimensions. We found that the local rank correlation closely corresponds to our visual interpretation of the quality of the output in benchmark examples.
We also demonstrated that the local rank correlation can be applied to the problems of estimating the intrinsic dimensionality of the original data, and selecting appropriate values for  tuning parameters.

\bibliographystyle{apa}
% This specifies the location of the file containing the bibliographic information.  
% It assumes you're using BibTeX (if not, why not?).
\cleardoublepage % This is needed if the book class is used, to place the anchor in the correct page,
                 % because the bibliography will start on its own page.
                 % Use \clearpage instead if the document class uses the "oneside" argument
\phantomsection  % With hyperref package, enables hyperlinking from the table of contents to bibliography             
% The following statement causes the title "References" to be used for the bibliography section:

% Add the References to the Table of Contents
\addcontentsline{toc}{chapter}{\textbf{References}}

\bibliography{lsp}
% Tip 5: You can create multiple .bib files to organize your references. 
% Just list them all in the \bibliogaphy command, separated by commas (no spaces).

% The following statement causes the specified references to be added to the bibliography% even if they were not 
% cited in the text. The asterisk is a wildcard that causes all entries in the bibliographic database to be included (optional).
\nocite{*}

\end{document}